\documentclass[amsmath,amssymb, superscriptaddress, reprint]{revtex4-2}
\pdfoutput=1
\usepackage{hyperref}
\usepackage{graphicx}
\usepackage{dcolumn}
\usepackage{bm}

\usepackage[utf8]{inputenc}
\usepackage[T1]{fontenc}

\newcommand{\GeSn}{Ge\textsubscript{1-x}Sn\textsubscript{x}}
\newcommand{\cmi}{cm\textsuperscript{-1}}

\begin{document}
	\title{Thermal expansion and temperature dependence of Raman modes in epitaxial layers of Ge and \GeSn}
	
	\author{Agnieszka Anna Corley-Wiciak}
	\affiliation{IHP – Leibniz-Institut für innovative Mikroelektronik, Im Technologiepark 25, 15236 Frankfurt (Oder), Germany}
	\affiliation{RWTH Aachen University, 52056 Aachen, Germany}
	\author{Diana Ryzhak}
	\affiliation{IHP – Leibniz-Institut für innovative Mikroelektronik, Im Technologiepark 25, 15236 Frankfurt (Oder), Germany}
	\author{Marvin Hartwig Zoellner}
	\affiliation{IHP – Leibniz-Institut für innovative Mikroelektronik, Im Technologiepark 25, 15236 Frankfurt (Oder), Germany}
	\author{Costanza Lucia Manganelli}
	\affiliation{IHP – Leibniz-Institut für innovative Mikroelektronik, Im Technologiepark 25, 15236 Frankfurt (Oder), Germany}
	\author{Omar Concepción}
	\affiliation{Peter Grünberg Institute 9 (PGI-9) and JARA-Fundamentals of Future Information Technologies, Forschungszentrum Jülich, 52428 Jülich, Germany}
	\author{Oliver Skibitzki}
	\affiliation{IHP – Leibniz-Institut für innovative Mikroelektronik, Im Technologiepark 25, 15236 Frankfurt (Oder), Germany}
	\author{Detlev Grützmacher}
	\affiliation{Peter Grünberg Institute 9 (PGI-9) and JARA-Fundamentals of Future Information Technologies, Forschungszentrum Jülich, 52428 Jülich, Germany}
	\affiliation{RWTH Aachen University, 52056 Aachen, Germany}
	\author{Dan Buca}
	\affiliation{Peter Grünberg Institute 9 (PGI-9) and JARA-Fundamentals of Future Information Technologies, Forschungszentrum Jülich, 52428 Jülich, Germany}
	\author{Giovanni Capellini}
	\affiliation{IHP – Leibniz-Institut für innovative Mikroelektronik, Im Technologiepark 25, 15236 Frankfurt (Oder), Germany}
	\affiliation{Dipartimento di Scienze, Università Roma Tre, V.le G. Marconi 446, 00146 Roma, Italy}
	\author{Davide Spirito}
	\email{spirito@ihp-microelectronics.com}
	\affiliation{IHP – Leibniz-Institut für innovative Mikroelektronik, Im Technologiepark 25, 15236 Frankfurt (Oder), Germany}

	
	\begin{abstract}
	Temperature dependence of vibrational modes in semiconductors depends on lattice thermal expansion and anharmonic phonon-phonon scattering. Evaluating the two contributions from experimental data is not straightforward, especially for epitaxial layers that present mechanical deformation and anisotropic lattice expansion. In this work, a temperature-dependent Raman study in epitaxial Ge and \GeSn~layers is presented.
	A model is introduced for the Raman mode energy shift as a function of temperature, comprising thermal expansion of the strained lattice and anharmonic corrections. With support of x-ray diffraction, the model is calibrated on experimental data of epitaxial Ge grown on Si and \GeSn~grown on Ge/Si, finding that the main difference between bulk and epitaxial layers is related to the anisotropic lattice expansion.
	The phonon anharmonicity and other parameters do not depend on dislocation defect density (in the range $7\cdot10^6$ – $4\cdot10^8$ cm\textsuperscript{-2}) nor on alloy composition in the range 5-14 at.\%.
	The strain-shift coefficient for the main model of Ge and for the Ge-Ge vibrational mode of \GeSn~is weakly dependent on temperature and is around -500 \cmi. In \GeSn, the composition-shift coefficient amounts to -100 \cmi, independent of temperature and strain.

	\end{abstract}
\maketitle

\section{Introduction}
Temperature dependence of vibrational modes in solids is driven by anharmonicity of the interatomic potential, which induces a shift of the phonon energy as a function of temperature, via two mechanisms: the direct (explicit) effect of high-order terms of the potential and the thermal lattice expansion\cite{grimvall_thermophysical_1999}. Anharmonicity is relevant in fundamental studies of electron-phonon interaction and phonon scattering associated to heat transport\cite{grimvall_thermophysical_1999}. On the practical side, phonon energy is measured by Raman spectroscopy, a technique of choice to investigate materials and microdevices with a fast turnaround and a non-destructive method. Raman spectroscopy can measure strain and composition of semiconductor alloys\cite{beechem_simultaneous_2007}, or be employed as a thermometry method\cite{sandell_thermoreflectance_2020,spirito_thermoelectric_2021}. All these quantities (temperature, strain, composition, anharmonicity) shift the modes’ energy with comparable order of magnitude (a few \cmi), which makes a quantitative assessment of anharmonic contributions difficult.

These characterizations are especially relevant for applications in optoelectronic devices such as lasers, whose performance depends critically on the ability to control the band structure through alloying and strain, operating over a broad range of temperature. A reliable method is especially required to support the development of group-IV, CMOS-compatible light sources\cite{marzban_strain_2023, buca_room_2022,armand_pilon_investigation_2022, moutanabbir_monolithic_2021, wang_gesnoi_2021, chretien_gesn_2019}. Additionally, thermoelectrics with group-IV alloys\cite{spirito_thermoelectric_2021, maeda_high_2022,kurosawa_new_2018} requires the assessment of phonon scattering mechanisms, which depend on the anharmonicity\cite{toberer_phonon_2011}. 

Such devices typically comprise heterostructures of Si, Ge, Sn, and their alloys, which are grown epitaxially onto Si and Ge substrates\cite{grutzmacher_sigesn_2023}. The thermal budget and mechanical constraints during deposition and processing results in mechanical deformation in the heterostructure and in biaxial strain of the epilayers, whose thermal expansion differs from a bulk material\cite{capellini_high_2012}. The phonons in strained layers shift with respect to the bulk, because the elastic constants depend on the deformation\cite{anastassakis_phonons_1998}. In the practice of Raman spectroscopy, strain $\varepsilon$ and composition $x$ are evaluated assuming a linear shift of the phonon energy $\omega=\omega_0 + ax + b\varepsilon$ (where $\omega_0$ is the energy for the unstrained material). The coefficients $a,b$ are obtained from calibration samples whose composition and lattice deformation is known thanks to independent measurements, e.g. x-ray diffraction (XRD)\cite{manganelli_temperature_2020, suess_power-dependent_2014, gassenq_raman_2017,vasin_structural_2018}. The temperature dependence of these modes' energy must be analyzed in the context of the constrained expansion.

All the features of temperature-dependent Raman scattering are also inherently correlated to the determination of phonon scattering mechanisms, especially in alloys. As mentioned above, this is relevant for several applications, and a clear understanding of these mechanisms can elucidate the role of alloy order/disorder configuration. Recent results show that \GeSn~ and silicon-germanium-tin can have peculiar configurations different from the random alloy ordering, depending on Sn content and thermodynamic conditions \cite{corley-wiciak_local_2023,cao_short-range_2020,jin_role_2023}. Here temperature-dependent Raman measurements will provide useful insight. Additionally, lab-scale Raman spectroscopy is a preferred alternative to methods that require large-scale facilities or destructive sample preparation, such as x-ray absorption fine structure (EXAFS)\cite{lentz_local_2023} or atomic probe tomography (APT)\cite{liu_3d_2022}.

Considering these practical and theoretical interests, in this work we aim to establish a method for extracting relevant parameters related to the anharmonicity of the phonons, together with $a,b$, from temperature-dependent Raman spectra. As model systems, and for the application interest, we will consider Ge and \GeSn~ epitaxial layers grown by chemical vapor deposition (CVD), of quality suitable for optoelectronics applications.

\section{Theory}
We study the effect of thermal expansion on the energy of the modes $\omega_E(T)$ with the equation \cite{grimvall_thermophysical_1999, ritz_thermal_2019}
\begin{equation}\label{eq:dweps_gen}
	d\omega_E = -\omega_E\sum_{ij} \gamma_{ij} d\xi_{ij}, \quad \gamma_{ij} = -\frac{1}{\omega_E}\frac{\partial\omega_E}{\partial\xi_{ij}}
\end{equation}
where $\gamma_{ij}$ is the (tensor) Grünheisen parameter and $\xi_{ij}$ is the deformation tensor of the unit cell by effect of temperature or external forces. Introducing the coefficient of thermal expansion (CTE) $\alpha_{ij}=({\partial \xi_{ij}}/{\partial T})$ and integrating the previous equation from a reference temperature $T_0$,  we obtain
\begin{equation}\label{eq:integ_wgen}
	\omega_{E}(T;\omega_0, \alpha, \gamma_{ij}) =   \omega_0\exp\left[-\sum_{ij}\gamma_{ij}\int_{T_0}^{T}\alpha_{ij}(t)dt\right].
\end{equation}

In the case of a cubic material, such as bulk Ge, that is free to expand and contract under the effect of temperature, the tensors are proportional to the identity matrix, e.g. CTE are diagonal with $\alpha_{11}=\alpha_{22}=\alpha_{33}=\alpha_0$. Similarly, the Grünheisen parameter is $\gamma_0$, and there is a single lattice parameter $a_0$, with $\alpha_0=(\partial a_0/\partial T)/a_0$. In contrast, for the case of epitaxial layers, such as Ge grown on Si, the thermal expansion is constrained by the substrate. In this case, the epilayer is biaxially strained, as its lattice parameters and CTE are different from the substrate. Thus, the epilayer features a tetragonal lattice with parameters $a_\parallel, a_\perp$ for the directions parallel and perpendicular to the interface, respectively. Similarly, the tensor quantities ($\gamma$, $\alpha$, $\varepsilon$) are diagonal in the reference frame aligned with the crystal axes, e.g., $\alpha_{11}=\alpha_{22}=\alpha_\parallel,\, \alpha_{33}=\alpha_\perp$. The strained lattice parameters are given by $a_j = a_0(1+\varepsilon_j)$, with $j=\parallel,\perp$. With these definitions, the CTE becomes
\begin{eqnarray}\label{eq:aj_eps}
	\alpha_j =&& \frac{1}{a_j}\frac{\partial a_j}{\partial T} =\nonumber\\
	&&\frac{1}{a_0 (1+\varepsilon_j)}\frac{\partial a_0(1+\varepsilon_j)}{\partial T} = \alpha_0 + \frac{\partial \ln(1+\varepsilon_j)}{\partial T}.
\end{eqnarray}

The parallel expansion is constrained to that of the substrate $\alpha_S$, i.e. $\alpha_\parallel=\alpha_S$. Additionally, the elastic properties of the epilayer tends to keep its volume constant, yielding the constraint $\varepsilon_\perp=-K\varepsilon_\parallel$, with $K=-2C_{12}/C_{11}=2\nu/(1-\nu)$, where $C_{ij}$ are the stiffness constants referred to the crystal axes and $\nu$ the Poisson's ratio\cite{anastassakis_phonons_1998}. 

Integration of Eq. \ref{eq:aj_eps}~ between $T_0$ and $T$, using the elastic constraints, gives the thermal expansion for several quantities of interest, reported below for convenience:

\begin{subequations}
	\label{eq:aepi}
	\begin{equation}
		\frac{a_\parallel(T)}{a_\parallel(T_0)} = \exp\left[\int_{T_0}^T\alpha_S(t)dt\right]
	\end{equation}
	\begin{equation}
		\frac{a_\perp(T)}{a_\perp(T_0)}= \exp\left[\int_{T_0}^T\left((1+K)\alpha_0(t) - K\alpha_S(t)\right)dt\right]
	\end{equation}
	\begin{equation}
		\frac{1+\varepsilon_\parallel(T)}{1+\varepsilon_\parallel(T_0)} = \exp\left[\int_{T_0}^T(\alpha_S(t)-\alpha_0(t))dt\right] 
	\end{equation}
\end{subequations}

The Raman mode energy dependence on temperature and strain can be obtained from Eq.\ref{eq:dweps_gen}. At fixed temperature $T$, the effect of strain is
\begin{eqnarray}\label{eq:strain_only}
	\frac{\omega_E(T, \varepsilon_\parallel)}{\omega_E(T, \varepsilon_\parallel=0)} = \frac{\omega^{tetra}(T)}{\omega^{cubic}(T)} &=&  \exp\left[-(2\gamma_\parallel -K\gamma_\perp) \varepsilon_\parallel\right]\nonumber\\
	&\approx& 1+\beta\varepsilon_\parallel
\end{eqnarray}
where the symbols $\omega^{tetra}$ and $\omega^{cubic}$ are introduced for the strained and unstrained case, respectively, and $\beta=-(2\gamma_\parallel -K\gamma_\perp)$.
The last equality, valid for small strain (e.g., thermal strain is ~10\textsuperscript{-3}), can be compared with the commonly assumed linear relation $\omega(\varepsilon_\parallel)-\omega(\varepsilon_\parallel=0)=b\varepsilon_\parallel$, with $b(T) = \beta\cdot\omega^{cubic}(T)$. For Ge, assuming $\gamma_\parallel=\gamma_\perp=\gamma_0\approx 1.3$ and $\omega_0 \approx300\,\textrm{cm}{^{-1}}$, we find $b\approx -500\,\textrm{cm}{^{-1}}$, comparable with reported values (see Ref. \onlinecite{manganelli_temperature_2020} and Refs. therein). $b$ depends on the temperature as the phonon energy in an unstrained material $\omega^{cubic}(T)$.

The temperature dependence $\omega_E(T)$ is obtained from Eq. \ref{eq:integ_wgen} knowing the $\alpha_j$s and $\gamma_j$s as
\begin{equation}\label{eq:dwdt_int}
	\frac{\omega_E(T)}{\omega_E(T_0)} = \exp\left[-\int_{T_0}^{T}\left( 2\gamma_\parallel\alpha_\parallel(t) + \gamma_\perp\alpha_\perp(t)\right)dt\right].
\end{equation}
The $\alpha_j(T)$ can be measured by XRD or obtained from known CTE. Vice versa, the measurement of $\omega_E(T)$ allows to extract the values of $\gamma_j$, provided the knowledge of CTE.

An alternative formulation of Eq. \ref{eq:dwdt_int} as explicit function of the biaxial strain $\varepsilon_\parallel$ is
\begin{eqnarray}\label{eq:strain_temperature}
	\frac{\omega_E(T, \varepsilon_\parallel)}{\omega_E(T_0, \varepsilon_\parallel=0)} &=&
	\exp\left[-(2\gamma_\parallel+\gamma_\perp)\int_{T_0}^{T} \alpha_0(t) dt\right]\nonumber\\
	&&\times\exp\left[\beta\, \varepsilon_\parallel(T)\right].
\end{eqnarray}

In Eq. \ref{eq:dwdt_int}, the anisotropic lattice expansion is used, while in Eq. \ref{eq:strain_temperature} the strain appears explicitly. Note that the energy scale is different: in Eq. \ref{eq:strain_temperature}, it is the energy at $T_0$ of an unstrained, cubic crystal  $\omega_0^{cubic}$, while in Eq. \ref{eq:dwdt_int}, it is the energy at $T_0$ for the tetragonal crystal ($\omega_0^{tetra}$). 

The explicit anharmonicity adds a correction $\Delta\omega_A(T)$ to $\omega_E(T)$ arising from high-order terms in the atomic potential at fixed volume. $\Delta\omega_A$ depends on the magnitude of the anharmonic interaction and the phonon density of states\cite{menendez_temperature_1984}. Rather than a calculation over the phonon branches, we considered a simplified model derived from Klemen's model with two terms (3- and 4-phonon interaction) and a simplified density of states\cite{menendez_temperature_1984,liu_temperature_2020}:
\begin{eqnarray}
	\label{eq:anh}	
	\Delta\omega_A (T; \omega_0, A_3, A_4) =&& -A_3\left[1+\frac{1}{e^{0.35x}-1} +\frac{1}{e^{0.65x}-1}\right]\nonumber\\
	&&- A_4\left[ 1+\frac{3}{e^{x/3}-1} \frac{1}{(e^{x/3}-1)^2} \right]
\end{eqnarray}
with $x=\hbar\omega_0/(k_BT)$. These terms account for the loss of energy of the phonon to other phonons; the coefficient $A_3$ is positive. Often, the 4-phonon term is ignored ($A_4\approx0$)\cite{liu_temperature_2020}. The complete formula is
\begin{equation}\label{eq:total}
	\omega(T) = \omega_{E}(T;\alpha, \omega_0, \gamma) + \Delta\omega_A (T; \omega_0, A_3, A_4).
\end{equation}

\begin{figure}
	\includegraphics{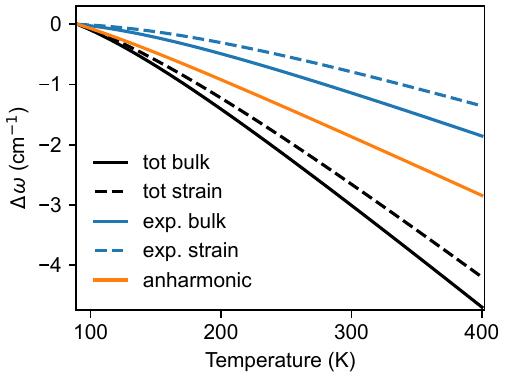}
	\caption{Calculated temperature dependence of the energy shift $\Delta\omega=\omega(T)-\omega(90 K)$ for bulk Ge and a strained Ge/Si layer. Expansion (label ``exp.'') and anharmonic terms, and their sum (expansion+anharmonic, label ``tot'') are reported. The anharmonic term (orange solid line) is the same for both Ge and Ge/Si. For the expansion term (and thus the total), the tensile strain in Ge/Si result (dashed lines) gives a shift of the energy. The parameters for the calculation are $\omega^{cubic}_0=304$ \cmi, $\gamma_0=\gamma_\parallel=\gamma_\perp=1.3$, $K=0.75$, $A_3=1$ \cmi, $A_4=0$, $\varepsilon(T_0)=0.2\%$, and $\alpha_{Ge}$ and $\alpha_{Si}$ from Ref. \onlinecite{roucka_thermal_2010}.}
	\label{fig:teo}
\end{figure}

Fig. \ref{fig:teo} shows the shift $\Delta\omega(T)=\omega(T)-\omega(T_0)$ calculated with Eq. \ref{eq:total} for bulk Ge and a Ge/Si heterostructure, as well as the expansion $\omega_E(T)-\omega_E(T_0)$ and anharmonic $\Delta\omega_A(T)-\Delta\omega_A(T_0)$ terms. Anharmonicity and thermal expansion have comparable effect, and the epilayer case also shows a non-negligible shift. For alloys such \GeSn, the effect of composition will be included, as discussed below. 

\section{Results and discussion}

The model must then be calibrated using samples with known strain and composition, to derive the relevant parameters such as $\gamma_j$s and $A_i$s. Therefore, we investigated several samples by Raman spectroscopy and XRD. Raman spectra as a function of temperature were acquired with excitation of 532 nm and a spectral resolution of around 0.7 \cmi. A liquid nitrogen cryostat controlled the temperature in the range 90-400 K. A selection of experimental spectra is shown in the Appendix, Fig. \ref{fig:ramanspectra}.

The lattice parameters as a function of temperature were measured by high-resolution XRD reciprocal space mapping. The tool featured a 9 kW rotating Cu Anode using line-focus geometry and was setup with a Ge(400)x2 channel cut monochromator. The sample environment was adjusted with a DCS 500 cooling stage enabling vacuum below 10\textsuperscript{-1} mbar and temperatures down to 90K. Temperature was calibrated with known Si and Ge wafers. The 004 and 224 Bragg positions were extracted from specular and asymmetric reciprocal space maps (RSMs), respectively. To obtain the parallel $a_\parallel$ and perpendicular $a_\perp$ lattice constant a biaxial strain model was applied assuming a tetragonal distortion\cite{zaumseil_role_2012}. The lattice spacing $d_{hkl}$ from a reflection $hkl$ is given by
\begin{equation}
	\frac{1}{d_{hkl}^2} = \frac{h^2+k^2}{a_\parallel^2} + \frac{l^2}{a_\perp^2}.
\end{equation}
In this way, from $d_{004}$ $a_\perp$ is calculated, while $d_{224}$ allows to calculate also $a_\parallel$. An example of RSM at low temperature is shown in the Appendix, Fig. \ref{fig:xrdspectra}, together with specular rocking curves of selected samples.

\begin{figure}
	\includegraphics{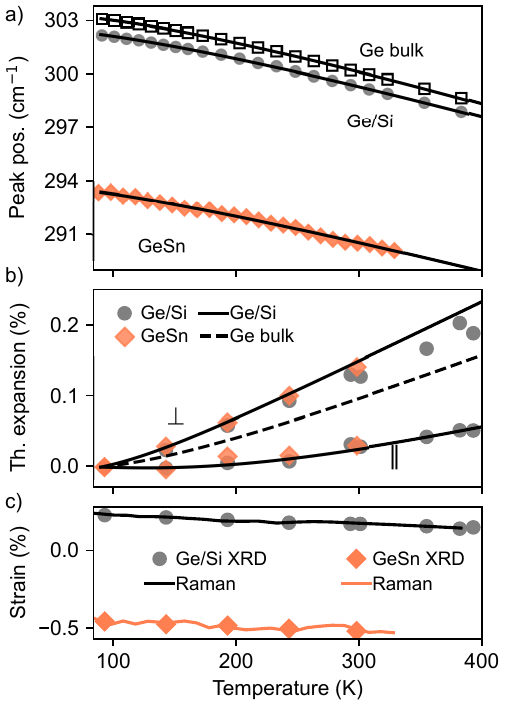}
	\caption{\label{fig:allfit}a) Peak position as a function of temperature for the main Raman peak of a a Ge bulk sample, a Ge/Si and a Ge\textsubscript{0.86}Sn\textsubscript{0.14}/Ge/Si heterostructure. Symbols are obtained as median of experimental data at 25 different positions, lines are the best fit described in the text. b) Thermal expansion in direction parallel ($\parallel$) and perpendicular ($\perp$) to the interface. Values are reported as $a_i(T)/a_i(T_0=90 K) -1$. Lines were calculated as discussed in the text. c) Biaxial strain as measured by XRD and as calculated from the Raman data for the Ge/Si and Ge\textsubscript{0.86}Sn\textsubscript{0.14}/Ge/Si samples in panel a.}
\end{figure}

As a first case, a wafer of pure Ge (``bulk Ge'') was studied, to evaluate $A_i$s and $\gamma_0$. Fig. \ref{fig:allfit}a reports the temperature dependence of the energy of main peak as empty squares. It was analyzed as a cubic, unstrained material to obtain the best fit values in Table \ref{tab:fit_results} and the relative line in the Fig. \ref{fig:allfit}a. The parameter $\gamma_0$ was compatible the reported value of 1.29\cite{roucka_thermal_2010}. $A_3$ is in the order of the \cmi~ and $A_4$ is much smaller (close to 0~\cmi).

Next, we considered a Ge/Si sample\cite{skibitzki_reduction_2020} with thickness 4.7~\textmu m and a Ge\textsubscript{0.86}Sn\textsubscript{0.14}(470 nm)/Ge(3.5 \textmu m)/Si sample\cite{spirito_thermoelectric_2021}. The heterostructures were grown on the (001) surface of Si substrates. For the \GeSn~case, we consider the peak of the main mode, assigned to Ge-Ge pair vibrations\cite{corley-wiciak_local_2023}. The values are reported as full diamonds in Fig. \ref{fig:allfit}a. The  temperature-dependent spectra for this sample are reported in the Appendix, Fig. \ref{fig:ramanspectra}b. For the Ge/Si, peak position is reported as full circles. The spectrum at 90 K is shown in Fig. \ref{fig:ramanspectra}a.

The Raman peaks shifts to lower energy as temperature increases, and are at lower energy with respect to Ge as result of strain and Sn content. The epitaxial strain in these samples enables the study of the anisotropy in $\alpha_j$s and $\gamma_j$s, using eqs. \ref{eq:dwdt_int} and \ref{eq:total}.

For this analysis, we measured the lattice thermal expansion via XRD for both parallel and perpendicular directions (reported as symbols in Fig. \ref{fig:allfit}b), as described before. These measurements were well reproduced by Eq. \ref{eq:aepi}, whose predicted values are reported in the figure as solid lines, while the dashed line refers to the lattice parameter of bulk Ge. The calculation used $\alpha_S=\alpha_{Si}$, $\alpha_0=\alpha_{Ge}$, $K=2\cdot0.373$,\cite{manganelli_temperature_2020} and reference values for the Si and Ge CTE\cite{roucka_thermal_2010, reeber_thermal_1996}, verifying the elastic constraint on the thermal expansion\cite{capellini_high_2012}. Both Ge/Si and \GeSn/Ge/Si follow the predicted trend. For the latter, the Ge layer was thin enough, so that the parallel expansion is dominated by the Si substrate. The good match between experimental and calculated values suggests also that the dependence of CTE and Poisson's ratio on Sn content is negligible at this composition.

\begin{table}
	\caption{\label{tab:fit_results} Best fit parameters for temperature-dependent Raman mode shift. Values with asterisks are calculated as $\omega_0^{cubic}=\omega_0^{tetra}\exp(-\beta\varepsilon_\parallel)$ with the strain $\varepsilon_\parallel$ measured by XRD and $b=\beta\cdot\omega_0^{cubic}$. Reference temperature is $T_0$=90 K.}
	\begin{ruledtabular}
		\begin{tabular}{llll}
			& Bulk Ge& Ge/Si&Ge\textsubscript{0.86}Sn\textsubscript{0.14}/Ge/Si\\ \hline
			$\omega_0^{cubic}$ (\cmi)& 304.2$\pm$0.1 & 304.7$\pm$0.3* &292.4$\pm$0.7*\\
			$\omega_0^{tetra}$ (\cmi)& - & 303.6$\pm$0.1 &294.8$\pm$0.2	\\
			$\gamma_0$& 1.2$\pm$0.2 & - &- \\
			$\gamma_\perp$&	- & 1.3 $\pm$ 0.1&1.3$\pm$0.3 \\
			$\gamma_\parallel$&	- & 1.3 $\pm$ 0.2 &1.3$\pm$0.3 \\
			$\beta=-2\gamma_\parallel+K\gamma_\perp$& - & -1.66$\pm$0.45* &  -1.64$\pm$0.53*\\
			$A_3$ (\cmi)&0.77$\pm$0.07 & 1.0$\pm$0.1&1.1$\pm$0.2 	\\
			$A_4$ (\cmi)&0.04$\pm$0.01& 0.018$\pm$0.009 &0.001$\pm$0.02  \\
			$b (90 K)$ (\cmi)&-&-510$\pm$140 &-470$\pm$150 \\
		\end{tabular}
	\end{ruledtabular}
\end{table}

For the epitaxial layers, the best fit values (Table \ref{tab:fit_results})  show that $\gamma_\parallel$ and $\gamma_\perp$ have similar values and matched with $\gamma_0$ of bulk Ge, within their uncertainty. Indeed, from a microscopic point of view, the $\gamma_j$s derive from the change of the interatomic potential with respect to the deformation of the crystal. Thus, a small anisotropy (as that in the epitaxial layers) will be negligible. The $A_3$ was slightly larger than bulk for both Ge/Si and \GeSn, and $A_4$ is negligible.

The strain in Ge/Si causes the shift with respect to bulk Ge as observed in Fig. \ref{fig:allfit}a. The strain as a function of temperature can be calculated from Eq. \ref{eq:strain_temperature} as
\begin{eqnarray}\label{eq:strain_es}
	\varepsilon_\parallel(T) = \frac{1}{\beta} \ln\left\{\frac{\omega(T)-\Delta\omega_A(T;\omega_0, A_4, A_4)}{\omega^{cubic}_0}\times\right.\nonumber\\
	\left.\times\exp\left[(2\gamma_\parallel+\gamma_\perp)\int_{T_0}^{T} \alpha_0(t) dt\right]\right\}.
\end{eqnarray}
In Fig. \ref{fig:allfit}c, we compare the strain measured by XRD and derived from Raman data, finding a good match and supporting the use of our model for strain estimation.

To validate the results, we have investigated a series of Ge/Si samples with different thickness and threading dislocation density (TDD) over more than one order of magnitude ($7\cdot10^{6}-4\cdot10^{8}$ cm\textsuperscript{-2})\cite{skibitzki_reduction_2020}. The $A_i$s and $\gamma_j$s were found to be independent of the sample, while $\omega_0^{tetra}$ changes slightly, matching XRD-measured strain.

\begin{figure}
	\includegraphics{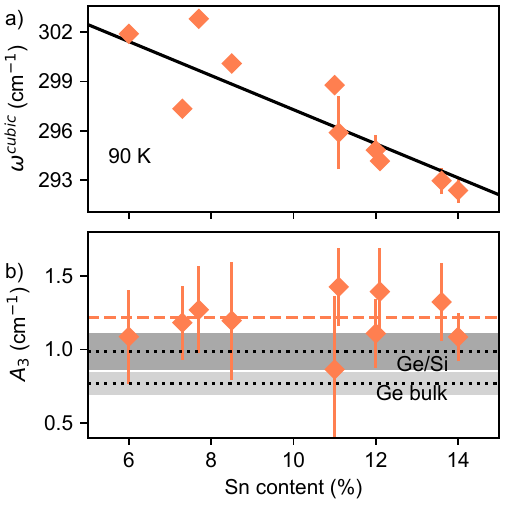}
	\caption{\label{fig:gesnresults}a) $\omega_0^{cubic}$ as a function of Sn concentration for \GeSn/Ge/Si layers. b) Explicit anharmonicity coefficient $A_3$. The dashed line is the weighted average of the data. The shaded areas mark the values for Ge/Si and Ge bulk. Mean values and errrobars are calculated from the fitting of several temperature-dependent spectra series at different position of each samples. For $\omega^{cubic}$, error bars are smaller than the symbols in some cases.}
\end{figure}

For \GeSn, $\omega_0^{cubic}$ shifts with respect to Ge because of alloying. Indeed, in a series of Ge\textsubscript{1-x}Sn\textsubscript{x}/Ge/Si samples ($x=5-14 \mathrm{at.\%}$), the major effect was a shift of the $\omega_0^{cubic}$ to lower energy as $x$ increases (Fig. \ref{fig:gesnresults}a). Using the XRD-measured strain and fitting the temperature-dependent Raman shift, we studied the effect of $x$ independent on the strain. As expected (see Ref. \onlinecite{corley-wiciak_local_2023} and Refs. therein), a linear trend was found. The linear regression  $\omega_0^{cubic}=ax+\omega^*$ yielded $a=-100\pm20$ \cmi, and $\omega^*=308\pm2$ \cmi at 90 K. The procedure was repeated in the range 90 K - 350 K. The coefficient $a(T)$ was found constant within the error, while $\omega^*(T)$ follows Eq. \ref{eq:integ_wgen} as an unstrained layer with $\omega_0=307\pm1$ \cmi~ and $\gamma_0=0.9\pm0.1$.

The other parameters were independent of composition within the error, supporting the robustness of values in Table \ref{tab:fit_results}. The coefficient $A_3$ (Fig. \ref{fig:gesnresults}b) was slightly larger than the Ge/Si and Ge cases, but without clear dependence on $x$.

A dependence of $A_3$ (i.e., of the anharmonic part of the interatomic potential) on the Sn content may be expected, given the higher mass of Sn with respect to Ge, despite the low content in the layer under investigation (below 14 at.\%). Nonetheless, recent theoretical studies \cite{cao_short-range_2020,jin_coexistence_2022} have highlighted that the silicon-germanium-tin alloy system may have multiple local-order configurations, depending on the composition and thermodynamics of the growth, and the local order affects the Raman scattering as well\cite{corley-wiciak_local_2023}. Thus, a straightforward dependence of the anharmonicity on composition may be difficult to predict. Indeed, the measured independence of $A_3$ from $x$ and TDD confirms the analysis based on the peak width in Ref. \onlinecite{bagchi_temperature_2011}, suggesting that more detailed investigation may be needed on a broader selection of samples.

The anharmonic interaction and its dependence on composition is also relevant for thermal conductivity\cite{chang_anharmoncity_2018}, with implications for thermoelectrics and optoelectronics device design. Here, several phonon scattering mechanisms are active, including alloy disorder\cite{gurunathan_alloy_2020}, and size of crystallites or microstructures\cite{khatami_lattice_2016}. The very weak effect of composition on $A_3$ suggest that alloy and other form of disorder are dominant in phonon scattering, and possibly exclude  phonon-phonon interaction as a mechanism for variation of thermal conductivity in alloy samples with comparable composition\cite{spirito_thermoelectric_2021,cheaito_experimental_2012,uchida_carrier_2019}.

Finally, the coefficient $\beta$ had similar value for Ge and \GeSn~independent of composition and TDD. The strain-shift coefficient $b=\beta\omega(T,\varepsilon=0)$ was $b\approx500$ \cmi, a value in line with literature (Ref. \onlinecite{manganelli_temperature_2020} and Refs. therein). For \GeSn, $b$ depends on composition as $b=\beta\omega(T, x, \varepsilon=0)=\beta\cdot(\omega^*(T)+ax)$. Nonetheless, its variation was inside the experimental error. Thus, in many cases it will be a sufficient approximation to use the same value for \GeSn~ independent of $x$, e.g. to measure the strain distribution in microdevices such as lasers operating at cryogenic temperature.

The samples investigated in this work belonged to a class of high-quality heterostructures with low Sn content, so that the comparable extracted values of the parameters, as shown in Table \ref{tab:fit_results} and Fig. \ref{fig:gesnresults} may not be surprising. However, as discussed before, the complexity of this alloy system requires a detailed and careful investigation. Future work can go in the direction of the ternary alloy silicon-germanium-tin, as well as to high-Sn content \GeSn~obtained, e.g., by MBE. Additional experimental techniques, such as EXAFS, APT, or neutron scattering, may give further insight on these issues.

\section{Conclusions}
In conclusion, we investigated temperature-dependent Raman shift in epitaxial Ge and \GeSn~using a model that includes anisotropic thermal expansion of epitaxial layers. This allows to separate the effects of strain from explicit anharmonicity and to determine the anisotropic Grünheisen parameters and anharmonicity strength.

The anharmonicity is rather independent of defect density and composition of the layers. The phonon energy shifts linearly with strain and composition, and we estimated the strain-shift coefficient $b\approx -500$ \cmi, that is weakly dependent on temperature and sample. For \GeSn~alloys, this model yields a strain-independent composition coefficient $a$, also weakly dependent on temperature, $a=-100\pm20$ \cmi.

The calculated parameters for Ge/Si and \GeSn/Ge/Si can be directly used for measurements of composition and strain for any temperature, or in thermometry experiments. The model can be applied in general to epitaxial materials to assess the relevance of anisotropic expansion and anharmonicity.

\begin{acknowledgments}
This work was partially funded the Deutsche Forschungsgemeinschaft under Project No. 299480227.
A. A. C.-W. and D. R. contributed equally to this work.
\end{acknowledgments}

\appendix
\section{Raman spectra and XRD data}

\begin{figure}
	\includegraphics{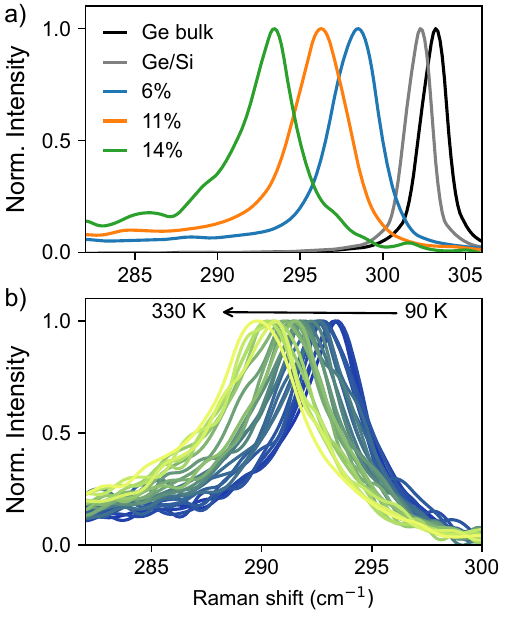}
	\caption{\label{fig:ramanspectra} a) Raman spectra of selected samples at 90 K. b) Spectra in the range 90-330 K for the a Ge\textsubscript{0.86}Sn\textsubscript{0.14}/Ge/Si sample. Curves are normalized for clarity.}
\end{figure}

Fig. \ref{fig:ramanspectra}a shows a selection of Raman spectra of Ge and \GeSn~ layers at low temperature. Panel b shows the temperature evolution for the case of the Ge\textsubscript{0.86}Sn\textsubscript{0.14}/Ge/Si sample.

\begin{figure}
	\includegraphics{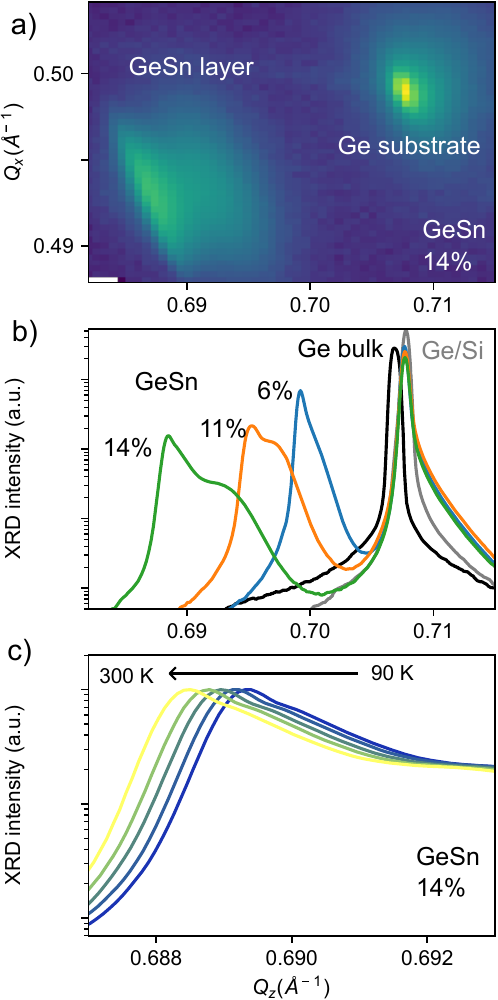}
	\caption{\label{fig:xrdspectra} a) Reciprocal space map of at 90 K of the Ge\textsubscript{0.86}Sn\textsubscript{0.14}/Ge/Si sample. b) Specular  curves as a function $Q_z$ for selected samples at room temperature, and c) as a function of temperature for the Ge\textsubscript{0.86}Sn\textsubscript{0.14}/Ge/Si sample.}
\end{figure}

Fig. \ref{fig:xrdspectra} shows the XRD data. A RSM measured at low temperature is shown in panel a for the Ge\textsubscript{0.86}Sn\textsubscript{0.14}/Ge/Si sample, and a series of temperature dependent curves for the same sample are in panel c. Panel b shows curves for the same samples as in  Fig. \ref{fig:ramanspectra}a.

\end{document}